\documentclass[aps,preprint,nofootinbib]{revtex4}
\usepackage{amsfonts}
\usepackage{amsmath}
\usepackage{amssymb}
\usepackage{graphicx}
\usepackage{epstopdf}
\usepackage{epsfig}
\newcommand{\be}{\begin{equation}}
\newcommand{\ee}{\end{equation}}
\newcommand{\bea}{\begin{eqnarray}}
\newcommand{\eea}{\end{eqnarray}}
\newcommand{\ba}{\begin{array}}
\newcommand{\ea}{\end{array}}

\newcommand{\eq}[1]{(\ref{#1})}

\begin{document}


\preprint{}
\title{Topological Order and Degenerate Singular Value Spectrum in Two-Dimensional Dimerized Quantum Heisenberg Model}

\vfill
\author{Ching-Yu Huang\footnote{896410093@ntnu.edu.tw} and  Feng-Li Lin\footnote{linfengli@phy.ntnu.edu.tw}  }
\affiliation{Department of Physics, National
Taiwan Normal University, Taipei,
116, Taiwan}

\vfill
\begin{abstract}
  We study the connection between topological order and degeneracy of the singular value spectrum by explicitly solving the two-dimensional dimerized quantum Heisenberg model in the form of tensor product state ansatz. Based on the ground state solution, we find non-zero topological entanglement entropy at the frustrated regime. It indicates a possible topological phase. Furthermore, we find that the singular value spectrum associated with each link in tensor product state is doubly degenerate only in this phase. Degeneracy of the singular value spectrum is robust against various types of perturbations, in accordance with our expectation for topological order. Our results support the connection among topological order, long range entanglement and the dominant degenerate singular values. In the context of tensor product state ansatz, the numerical evaluation of singular value spectrum costs far less computation power than the one for topological entanglement entropy. Our results provide a more viable way to numerically identify the topological order for the generic frustrated systems.

\end{abstract}

\maketitle

\*\\
 \section{Introduction}

Investigation of topological orders for spin models is under
intensive study in the recent years, partly due to its relation to
the ground state of high Tc superconductor, the so called spin
liquid phase \cite{LGW}, and partly because of the enlightenment of quantum
information to characterize the ground states by quantum
entanglement \cite{Osborne}. These studies lead to new classification schemes
\cite{wen1d,wen2d,classifyMPS,Pollmann09,Pollmann08,2dDEE} of the quantum
phases beyond the usual Landau-Ginzburg-Wilson paradigm.  The new quantum phases such as the spin liquids are called the topological phases and cannot be characterized by a local order
parameter. Instead, it could be characterized by the ground state degeneracy, quasiparticle
statistics, existence of edge states, topological entanglement entropy \cite{kiteal06,wen06}, and entanglement spectrum \cite{Haldane08}, i.e., eigen-spectrum of the reduced density matrix. The latter twos are motivated by the quantum information studies  on the entangled nature of ground state, which could be efficiently approximated by the projected entangled pair state ansatz \cite{PEPS}.  One typical example for all the above characteristics of topological ordered state is the fractional quantum Hall states \cite{wen90,Friedman08}.

   For one-dimensional (1D) spin chain the quantum fluctuation could be more significant
than in the higher dimensional ones so that it will melt the system into liquid, and
thus becoming topological. The simplest example of the ``topological
phase" is the Haldane phase. The term with the quotation remark is to remind that
the topological nature of the ground state is protected by some symmetry of the spin chain \cite{Pollmann09,Pollmann08}.  Once the symmetry is broken, the ground state wave function is equivalent to a trivial product state up to some local unitary transformation \cite{qsrg}. The local unitary transformations can remove the short range entanglement, i.e., entanglement between the neighboring sites. In contrast, the long range entanglement is the entanglement among the distant sites, and cannot be removed by the local unitary transformations. Due to its robustness against the local operations, the long range entanglement causes the topological order of the nontrivial ground state.

   On the other hand, for two-dimensional (2D) spin systems the topological phase is usually expected to exist in the frustrated systems since the large ground state degeneracy of the system is highly quantum and cannot have simple classical order. However,  suffering from the lack of the reliable numerical
tools in solving the 2D frustrated spin systems,  only very few topological phases have been identified, except for some particularly exactly solvable models \cite{toric,levinwen,kitaevm}.  The connection between the topological order and the quantum entanglement is less understood than in the 1D cases.

  Despite of the recent progress in  \cite{wen1d,wen2d,classifyMPS,Pollmann09,Pollmann08,2dDEE} on the general scheme of classifying the topological orders caused by the long range entanglement,  it is still not easy in practical to dynamically determine if exists a topological phase.  It is then desirable to find a more numerical viable characterization for the topological order. Since the topological order is believed to be linked to the long range entanglement, we think the degeneracy pattern of the entanglement spectrum can be such an index. The intuitive picture is that the entanglement spectrum is the Schmidt values of the bi-partition of the whole system \cite{Pollmann08,2dDEE}, so the dominant degenerate singular values  implies that the wave function is close to the maximally entangled state (Bell state). In this case, the two parts of the system is coherently long-range entangled.

The entanglement spectrum is easy to evaluate numerically for 1D
spin systems in the ansatz of matrix product states (MPS) because it
coincides with the Schmidt value spectrum of the bi-partition, and
can be obtained by performing the singular value decomposition (SVD)
for the matrix of the MPS. Indeed, the connection between
entanglement spectrum and topological order is established for the
MPS  numerically \cite{Pollmann08}. However, this is not the case
for the 2D tensor product state (TPS) because the left and right
states of the bi-partition are now the 1D spin chains \cite{2dDEE},
and the corresponding entanglement spectrum is not the same as the
singular value spectrum. The latter is obtained by merging two neighboring 1-site tensors in TPS and then performing the SVD on the merging tensor.  Despite
that, these twos should be related, especially for the
translationally invariant states. Therefore, if the entanglement
spectrum can be used to characterize the 2D topological order, so
does the singular value spectrum. Indeed, we will see this is the
case in this work.

The main motivation of this paper is to establish such an connection in 2D among topological order, long range entanglement and degenerate singular value spectrum by numerically studying a particular frustrated 2D spin model with its ground state in the form of the TPS ansatz. In the meantime we explore the properties of the singular value spectrum under various types of the local unitary transformations to test its robustness, which is the main characteristics of the topological order.

 The model we will study is the staggered dimer spin 1/2 model on the square lattice, and will be referred as the  J-J' model form now on. It is an interesting 2D spin model with various issues
about quantum phase transition. For the antiferromagnetic $J'$-bond regime, i.e., $J'/J>0$, this model has been studied by using  the perturbation theory \cite{pertu88,RSWT96}, exact diagonalization
\cite{CCM00}, the coupled cluster method \cite{CCM00}, iPEPS \cite{JJPEPS} and
quantum Monte Carlo \cite{QMC08,QMC09}. In this regime, there  exists a quantum phase transition by tuning $J'$ with a critical point at $J'/J \approx 2.51$ \cite{pertu88}, separating the classical N\'{e}el ordered phase and a finite-gap disordered phase  \cite{QMC08,QMC09}.

 In this paper we are more interested in the regime of the ferromagnetic $J'$-bond, i.e.,
$J'/J<0$ because the plaquettes are now frustrated. The quantum Monte Carlo
method cannot be used for the frustrated system because of the sign
problem. Instead, some mean-field based approximation methods such
as the renormalized spin wave theory \cite{RSWT96}, exact diagonalization \cite{CCM00} and coupled cluster method \cite{CCM00} have been used to study this
regime. These studies found a phase transition at $J'/J=-1/3$
separating the N\'{e}el phase from a helical phase for classical
spins, and the maximal frustration occurs around $J'\approx1$
region. However, due to the frustrations we suspect that a quantum spin liquid phase with topological order may appear around the region of maximal frustration. We indeed find this is the case from our study.

   The paper is organized as follows. In the next section we describe the Iterative Projection method of solving the 2D TPS ground state \cite{Jiang}, in the literatures this method is also referred as the ``simplified update"  \cite{2dsolve}. This is the 2D generalization of  the infinite time evolving block decimation (iTEBD) \cite{itebd}.  We also review the method of the the tensor-network renormalization group (TRG) \cite{trg1,trg2}, which are used to find the overlap of the wave functions in a polynomially efficient way. Based on these, we can obtain the singular value spectrum and topological entanglement entropy, and also perform the quantum state renormalization. We also briefly review the above concepts. In section 3 we present the phase diagram of the J-J' model by evaluating the order parameter, entanglement entropy, dimer strength and the ground state energy. In section 4 we evaluate the topological entanglement entropy for the J-J' model and a toric code like state to characterize the topological phase. In section 5 we focus on various properties of the singular value spectrum and its degeneracy to establish its connection with the topological order. Finally, we conclude our paper in section 6.

 \section{The Model and the Method}

 The model we study in this paper is the staggered dimer model on square lattice, also known as the J-J' model. It is a quantum Heisenberg antiferromagnetic model with two
kinds of nearest-neighbor exchange couplings, and its Hamiltonian is
\begin{align}
H=J\sum_{<ij>}\vec{S_i}\cdot \vec{S_j}+J'\sum_{<ij>'}\vec{S_i} \cdot\vec{S_j}\;,
\end{align}
where the summations over $<ij>$ and $<ij>'$ represent sums
over the nearest-neighbor bonds as shown in Fig.1. Each square lattice
consists of three $J$ bonds (the thin ones) and one $J'$ bond (the thick one). We fix the  $J$ bonds
to be antiferromagnetic, i.e., $J>0$, and consider $-\infty < J'/J <\infty $
as the parameter of this model.

\begin{figure}[ht]
\center{\epsfig{figure=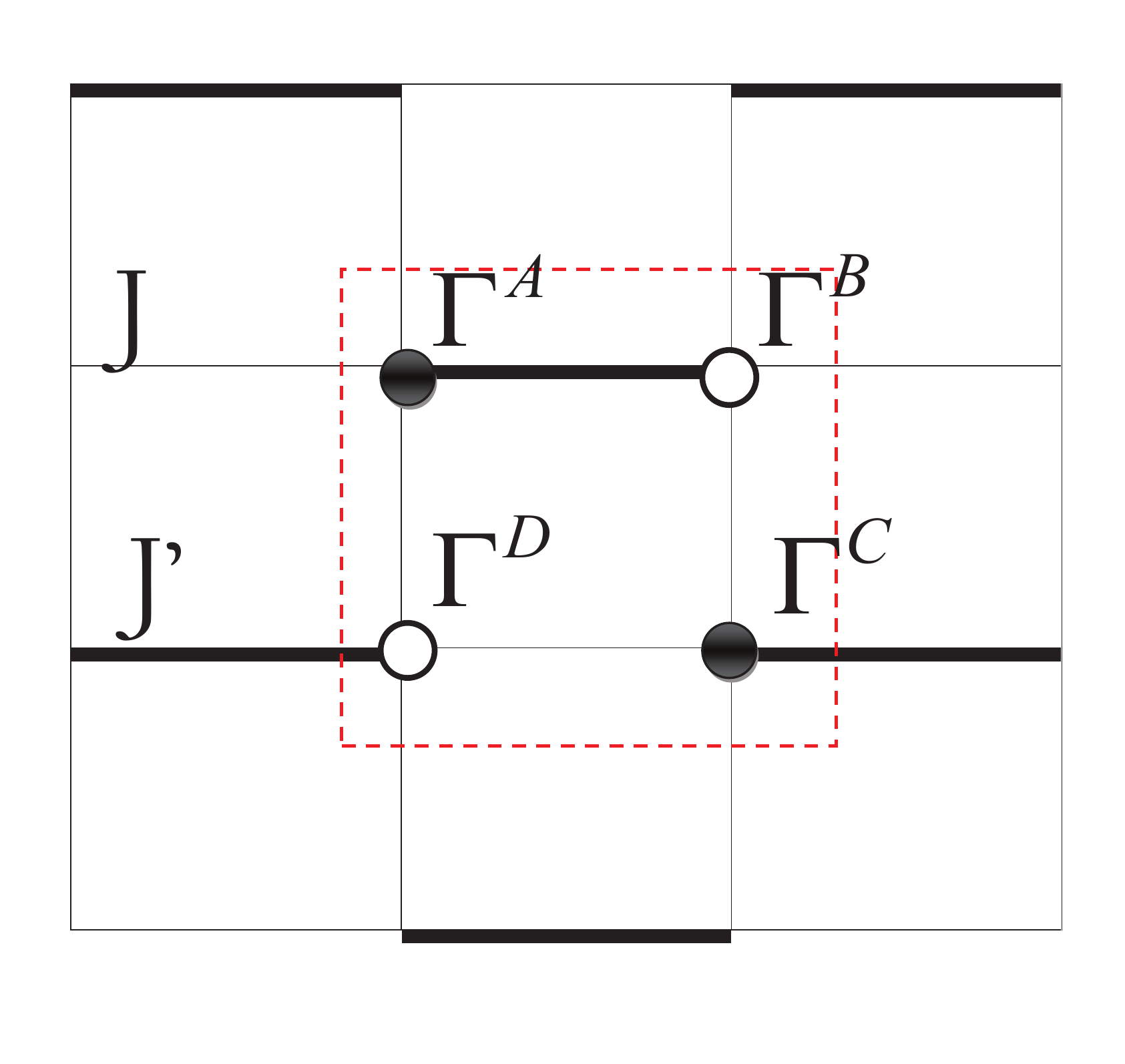,angle=0,width=8cm}} \caption{The
J-J' model on the 2D square lattice with two different nearest
neighboring bond couplings J and J' (thin and thick, respectively).
The red dash square is a $2\times2$ unit cell.} \label{model}
\end{figure}

\subsection{Solving the model with Iterative Projection and TRG}

   We will use the Iterative Projection algorithm \cite{Jiang} (a.k.a. simplified update \cite{2dsolve})  to find the translationally invariant ground state of the system in the form of the tensor product state ansatz
\begin{align} |\psi\rangle = \sum_{s_{1,1},s_{1,2},\ldots
,s_{N,N}} tTr[A^{[1,1]}(s_{1,1})A^{[1,2]}(s_{1,2})\ldots
A^{[N,N]}(s_{N,N})]|s_{1,1},s_{1,2},\ldots ,s_{N,N}\rangle \label{TPSg}
\end{align}
where $s_{i,j}$ labels the physical spin of dimension $d_s$ at site $(i,j)$ for
$i,j=1,\ldots , N$, $A^{[i,j]}_{r,d,l,u}(s_{i,j})$'s are rank-five tensors with the
bond indices $r,d,l,u=1,\cdots,\chi$, and  $tTr$  is to sum over all indices of tensors.
We call $\chi$ the bond dimension and $d_s$ the physical dimension.  This ansatz is able to handle an area-law by construction, but that doesn't mean that the true ground state of the system necessarily obeys it. Depending on the situation, we should tune $\chi$ to fully capture the essential feature of the ground state.

  The basic idea of Iterative Projection method is to evolve the system along the imaginary time in a very small time step so that the time evolution operator can be expanded as a sequence of two-site unitary operations through the Suzuki-Trotter decompositions.  For long enough time evolution, the system will go to the ground state we are solving for.  We further assume translational invariance of the ground state ansatz \eq{TPSg}, then we only need to update 4 tensors  as shown in Fig. 1 inside a unit cell for each time step.

   Based on the solved ground state from Iterative Projection method, we can calculate some quantities to characterize the phase diagrams. For the symmetry breaking phase we can evaluate the order parameters which are the vacuum expectation value (vev) of some physical operators. However, when we insert a physical operator at some particular site such that the translational invariance is lost, one needs some efficient method to contract the exponentially large number of bounds. Here we will adopt the TRG method \cite{trg1,trg2} to do that. The basic idea of TRG is to renormalize the tensors of the TPS by keeping the relevant entanglement when coarse graining. In this way, we can  reduce the size of the system while retaining the essential quantum correlations of the original ground state. Finally the whole system will be reduced to a unit cell with the renormalized tensor, we can then evaluate the vev's  faithfully enough within the renormalized unit cell. Besides, there are alternative ways to implement polynomially efficient evaluation of the vev. One interesting method is to contract the tensors in the 1D scheme vertically and horizontally \cite{Jordan}, and the other one is to use the Monte Carlo sampling to enhance the rate of contracting the tensors \cite{Wang}.

\subsection{Order parameters for N\'{e}el and dimer phases}

    For the J-J' model considered here, we will evaluate the N\'{e}el order parameter i.e., given by  $M^z_s=\frac{1}{M}\sum_{i=1}^N (-1)^i \langle\psi_g|S_i^z|\psi_g\rangle$ to characterize the N\'{e}el ordered phase. We also evaluate the spin-spin correlations and the dimerization
\cite{leung96,Singh99,Yu96}:
\begin{align}\label{Dx}
D_x=|\langle \vec{S}_{i,j} \cdot \vec{S}_{i+1,j} \rangle - \langle \vec{S}_{i+1,j} \cdot \vec{S}_{i+2,j} \rangle|
\end{align}
and
\begin{align}\label{Dy}
D_y=|\langle \vec{S}_{i,j}\cdot \vec{S}_{i+1,j} \rangle - \langle \vec{S}_{i,j} \cdot \vec{S}_{i,j+1} \rangle|
\end{align}
to characterize the dimmer strength of the  disordered ``dimer phase". Note that the Hamiltonian of the model does not possess the dimerized order by construction.   Besides, we also use TRG to evaluate the ground-state energy per site as $\frac{E}{N}=\langle\psi_g|H|\psi_g\rangle /N$.

\subsection{Characterizing topological phase by topological entanglement entropy}

   On the other hand, it is more difficult to characterize the topological phase since there is no symmetry breaking order parameter for it. On typical quantity for the 2D systems is the topological entanglement entropy \cite{kiteal06,wen06}. It is the sub-leading constant term in the entanglement entropy
\be
S_L =\alpha L -\gamma + {\cal O}(L^{-\nu})\;, \qquad  \nu>0\;,
\ee
where $L$ is the boundary size of the block for which the entanglement entropy is evaluated by tracing out the degrees of freedom outside it.

\begin{figure}[ht]
\center{\epsfig{figure=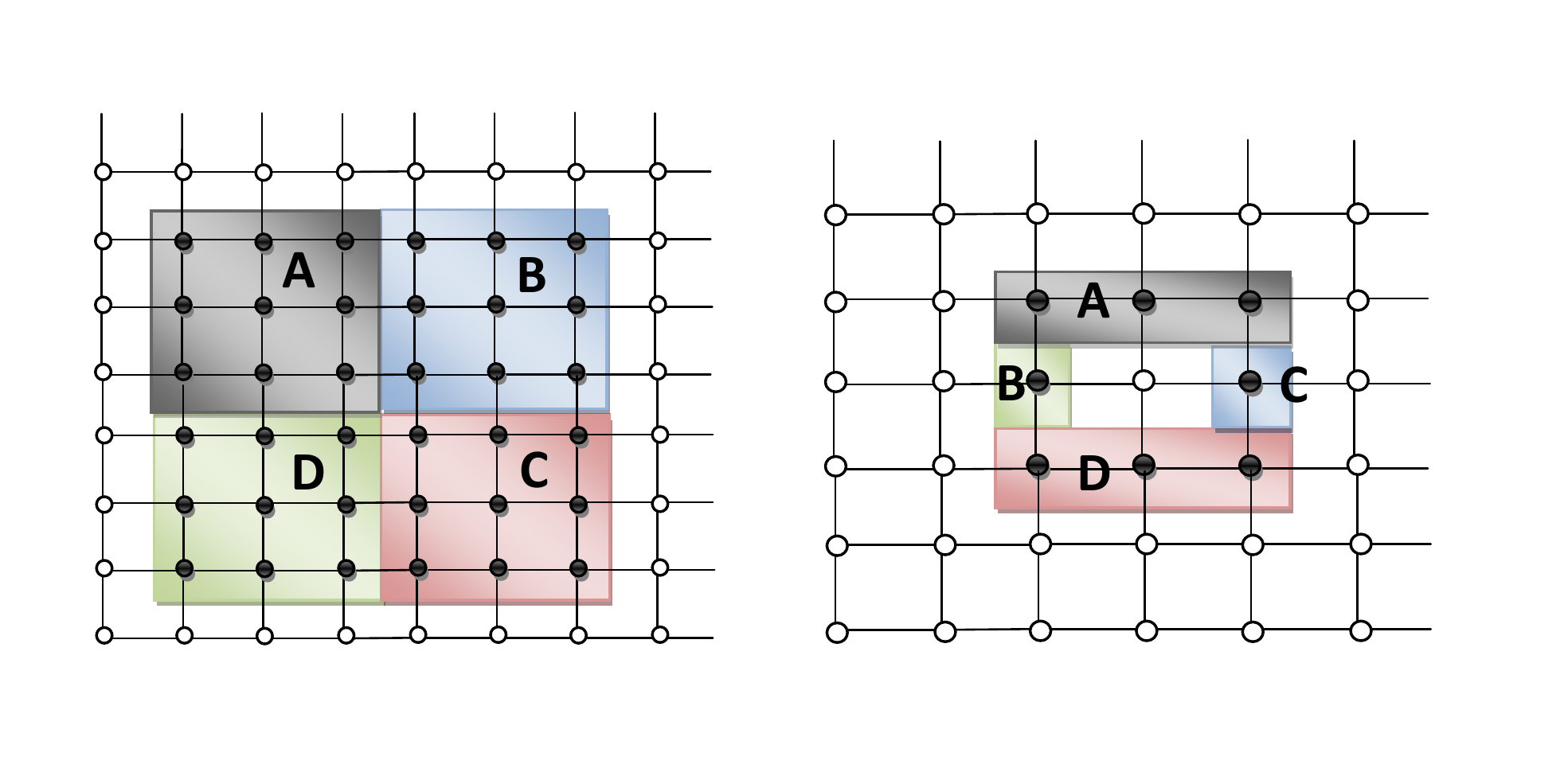,angle=0,width=13cm}}
\caption{Two schemes for evaluating the topological entanglement entropy.
(Left) Kitaev-Preskill's scheme for the square block with even number of sites on each side.  (Right) Levin-Wen's scheme for the square block with odd number of sites on each side.  }
\label{topocon}
\end{figure}

   Since the constant term $\gamma$ is topological and universal, we can extract it by appropriately subtracting the entanglement entropies of different blocks. There are two subtraction schemes, one is proposed by Kitaev and Preskill \cite{kiteal06}, and the other is by Levin and Wen \cite{wen06}. For the square lattice it is easier to implement the numerical evaluation of the entanglement entropy for a square or rectangular block. In this case, we find that it is convenient to adopt Kitaev-Preskill's scheme as shown in the left of Fig. \ref{topocon} for the square block with even number of sites on each side of the square block. The extraction for the topological entanglement entropy then goes as follows
\begin{align}
S_{topo}^{KP}=S_A+S_B+S_C+S_D-S_{AB}-S_{BC}-S_{CD}-S_{AD}+S_{ABCD}\;,
\end{align}
where $S_{AB..}$ denotes the von Neumann entropy of the density
matrix $\rho_{AB...}$ in the region $AB...\equiv A\cup B \cup ... $.

   As for the square block with odd number of sites on each side, it is convenient to adopt Levin-Wen's scheme as shown in the right of Fig. \ref{topocon}, and the extraction is
\begin{align}
S_{topo}^{LW}=S_{ABCD}-S_{ABD}-S_{ACD}+S_A+S_D\;.
\end{align}
In both cases, $S_{topo}=-\gamma<0$, which is related to total quantum dimension $D$ by $\gamma=Log D$ \cite{kiteal06,wen06}.

 \subsection{Characterizing topological phase by degeneracy of singular value spectrum}

    Another quantity in characterizing the topological phase is the degeneracy of the entanglement spectrum. It has been used to characterize the topological order for some quantum Hall states \cite{Haldane08}, and recently to characterize some symmetry-protected topological ordered Haldane phase in spin 1 chain \cite{Pollmann08}. The entanglement spectrum is refereed to the spectrum of the Schmidt values for the bi-partition of the whole system. By definition, these Schmidt values are the square roots of the eigenvalues of the reduced density matrix of either of the two partitions. Basically, the degeneracy of the entanglement spectrum implies the spins at different sites are almost in the maximally entangled state, hence it implies that the ground state has topological order.  In 1D, the entanglement spectrum is the same as the singular value spectrum of the matrix product state, and its power in characterizing the topological order is fully demonstrated \cite{wen1d,classifyMPS,Pollmann09,Pollmann08}. However, for general 2D spin system there is no consensus on the power of entanglement spectrum in characterizing the topological order, see \cite{2dDEE} for recent study on this issue \footnote{In \cite{2dDEE}, the entanglement spectrum of a bulk region is associated with a the boundary excitation spectrum. Their numerical study revealed that the boundary Hamiltonians become non-local for the topologically ordered states.}. Despite that, one may expect that it still works in 2D since the topological order is closely related to the long range entanglement characterized by the degeneracy of the entanglement spectrum.

    Unlike the 1D case, the 2D entanglement spectrum is not the same as the singular value spectrum of the 1-site tensor in TPS though these twos should be related, especially for translationally invariant states. Numerical task in evaluating the 2D entanglement spectrum is almost the same as for the topological entanglement entropy, and thus difficult. Instead, it is far more easier and straightforward to evaluate singular value spectrum for TPS by merging the tensors of neighboring sites along x- or y-direction and then doing SVD.  In this way, we can obtain four kinds of the singular value spectrum, which are associated with the four different links out of a site. For example, in the J-J' model there are three $J$-bonds and  one $J'$-bond, and they may have different singular spectrum. This can be done by just using Iterative Projection method without further invoking TRG.
    
      Besides, from the singular value spectrum one can straightforwardly evaluate the bipartite entanglement measure per length as well as the single-site von Neumann entropy (1-tangle). Both can be used to characterize the quantum critical point \cite{huang-lin}.

\subsection{Quantum state renormalization}

     Since the topological order is closely related to the long range entanglement, which however, is usually contaminated by the short range entanglement.  In order to make the characteristic of the topological order more explicit, one can implement the quantum state renormalization method \cite{qsrg,0qrg,1drg,2drg}, which can remove the short range entanglement by performing local unitary transformation.  The successive quantum state renormalization will then flow the original state to a simpler fixed-point state, which, however, has the same topological order of the original state. We can then classify the topologically ordered phases by the TPS's tensors of the fixed-point states.

     We now briefly describe the method as follows, and for more details see \cite{2drg}.  We first form a positive double tensor $\mathcal{T}$ by merging two layers of tensors $T$ and $T^{\dagger}$ with the physical indices contracted. Note that the resultant double tensor is constructed so that it is invariant under the local unitary transformation, and then we can spectrally decompose it again back into two tensors $\tilde{T}$ with physical indices. The spectral weights encode the relevance of the entangled components.  In this step we have removed the short range entanglement by the local unitary transformation taking the tensor $T$ into $\tilde{T}$ according to the significance of the spectral weights. Next, we will coarse grain the lattice labelled by the tensor $\tilde{T}$ by implementing one step of block decimation in the TRG method.

     After repeating the above steps the original TPS will then flow into a fixed-point state with all the short range entanglement removed. If we perform quantum state renormalization while keeping the symmetry or gauge symmetry of the lattice, the final fixed-point tensors can then be used to classify the topological phases according to the symmetry structure.

 \section{Numerical results for the phase diagram of J-J' model}
In this section, we will present the numerical results for the phase diagrams of the J-J' model of system size $N=2^8\times2^8$.

  We first evaluate the vev of staggered  magnetization $M_s^z$ in ground state as a function of $J'/J$ by using the TRG method, and the result is shown in Fig. 2.  In our calculation we consider the bond dimension up to $\chi=6$ and keep $D_{cut}\geq \chi ^2$ to ensure the accuracy of the TRG calculation. Here, $D_{cut}$ is the cutoff on the bond dimension of the merging lattice during the coarse graining in the TRG method. We find that the numerical results converge quickly along range entanglementady for $\chi=4$ and the numerical program works efficiently. Therefore, we will take $\chi=4$ and keep $D_{cut}=24$ in our numerical study in this paper.

\begin{figure}[ht]
\center{\epsfig{figure=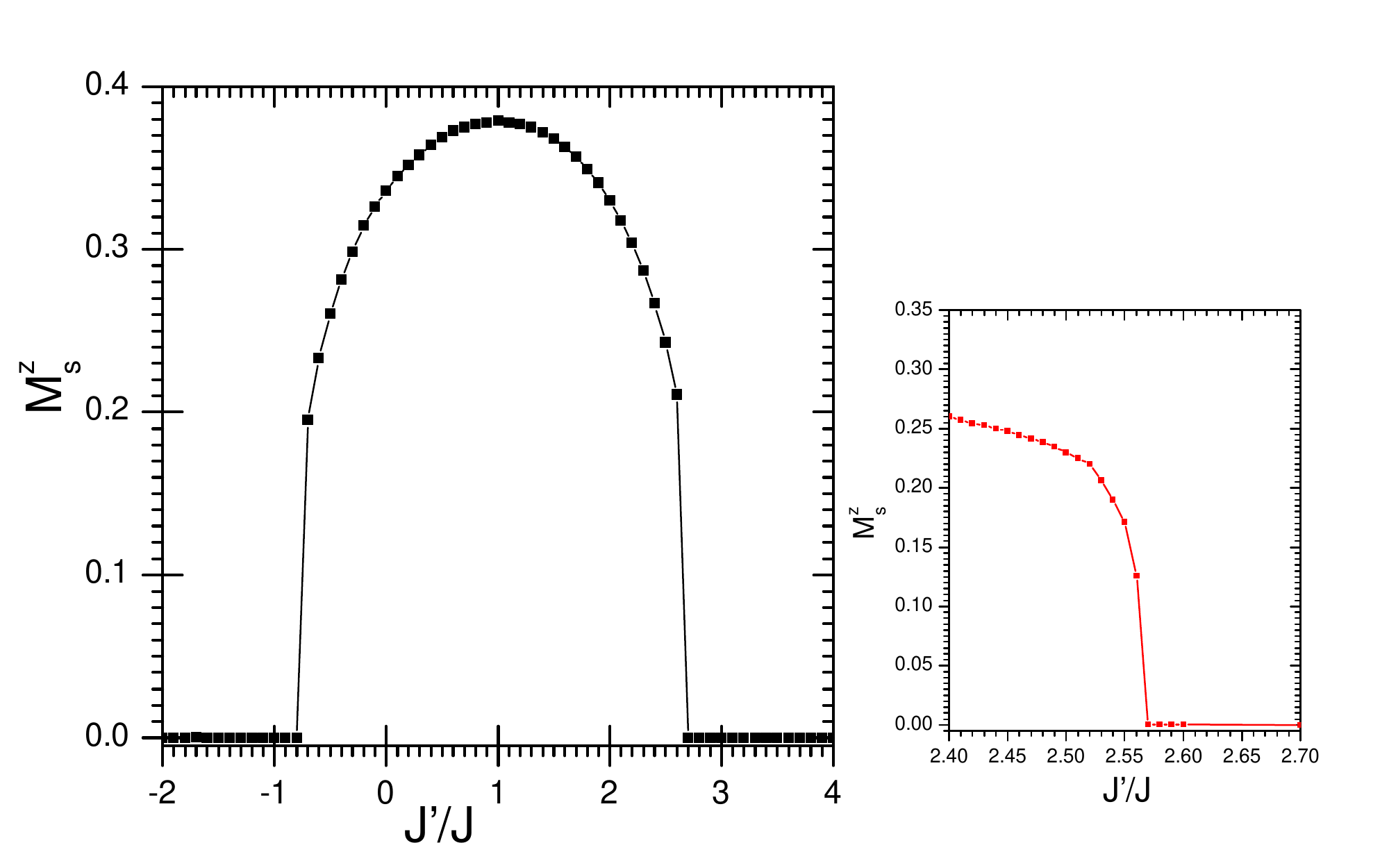,angle=0,width=12cm}} \caption{
Magnetization $\langle M^z_s \rangle$ v.s. $J'/J$ (Magnified the
regime around $J'/J\simeq 2.56$ in the right).  The ground state is
in the TPS ansatz with bond dimension $\chi=4$ and $D_{cut}= 24$ for
TRG in taking the vev. It indicates a second-order phase transition
from N\'{e}el to a disordered (dimerized) phase at $J'/J \approx
2.56$, and another quantum critical point at $J'/J \approx -0.54$
separating the N\'{e}el phase from a possible topological phase.}
\label{ms}
\end{figure}

 From Fig. \ref{ms}, it is observed that  $\langle M^z_s \rangle$ drops to zero both at $J'/J \approx 2.56$ and $J'/J \approx -0.54$. This indicates two quantum critical points at the corresponding critical values of $J'/J$.   It was known that a second-order transition from a N\'{e}el-ordered phase to a finite-gap disorder phase occurs at $J'/J \approx 2.51$, which is obtained from the unbiased quantum Monte Carlo simulation \cite{QMC08,QMC09}. Our result instead finds a critical point at $J'/J \approx 2.56$. The discrepancy could be due to the not large enough $\chi$ used in our numerical calculation. Furthermore, we fit the critical exponent $\beta$ for the magnetization, i.e.,  $M_s=A|J'-J_c'|^\beta$. Our result is $\beta\simeq 0.37691$, which is close to the exponents of the 3D classical Heisenberg $(O(3))$ model, i.e., $\beta\simeq 0.3639\pm0.0035$ \footnote{By increasing $\chi$ one may improve the above the results to be close to the results by the quantum Monte Carlo method, though it may require far large computational power. In this work, we are more interested in the identification of the topological phase, and will be satisfied by the above accuracy for the $O(3)$ phase transition.}.

\begin{figure}[ht]
\center{\epsfig{figure=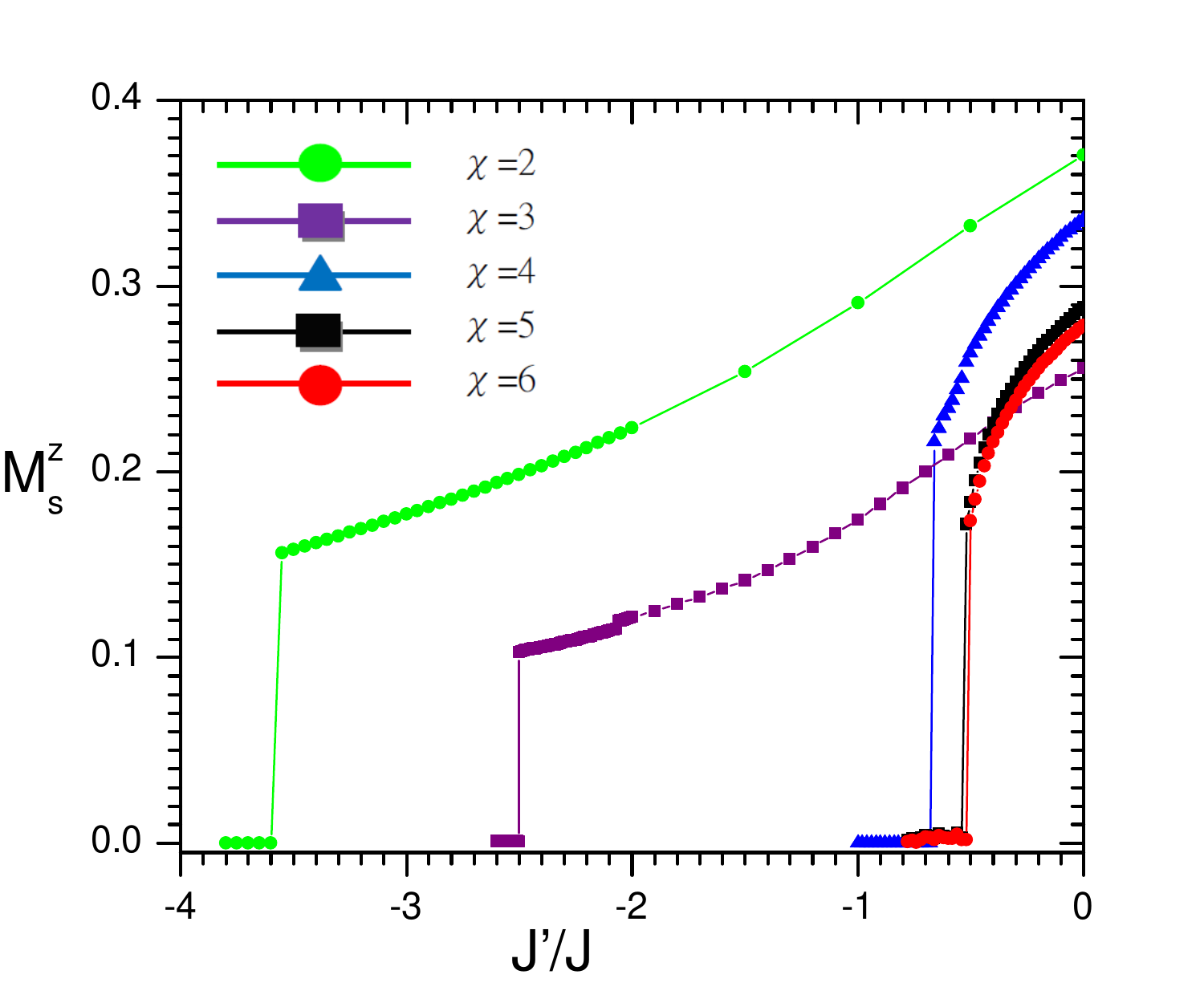,angle=0,width=10cm}}
\caption{$\langle M^z_s \rangle$ vs $J'/J$ for different bond
dimensions, the $\chi$'s. Larger $\chi$ yields a critical behavior
more like a second order phase transition, other than the first
order one. For $\chi=6$ it indicates a second order phase transition
at $J'/J\simeq -0.54$.} \label{msmi}
\end{figure}

  The other critical point at $J'/J \approx -0.54$ is frustrated since $J'<0$ and $J>0$, and there is possibility for the existence of topologically ordered phase due to the underlying degeneracy of the frustrated ground state. In Fig. \ref{msmi} we show how the critical behavior in this regime changes when we change the bond dimension.  Our result showing $\langle M^z_s \rangle=0$ for $J'/J < -0.54$ is consistent with this expectation for a topologically ordered phase.

\begin{figure}[ht]
\center{\epsfig{figure=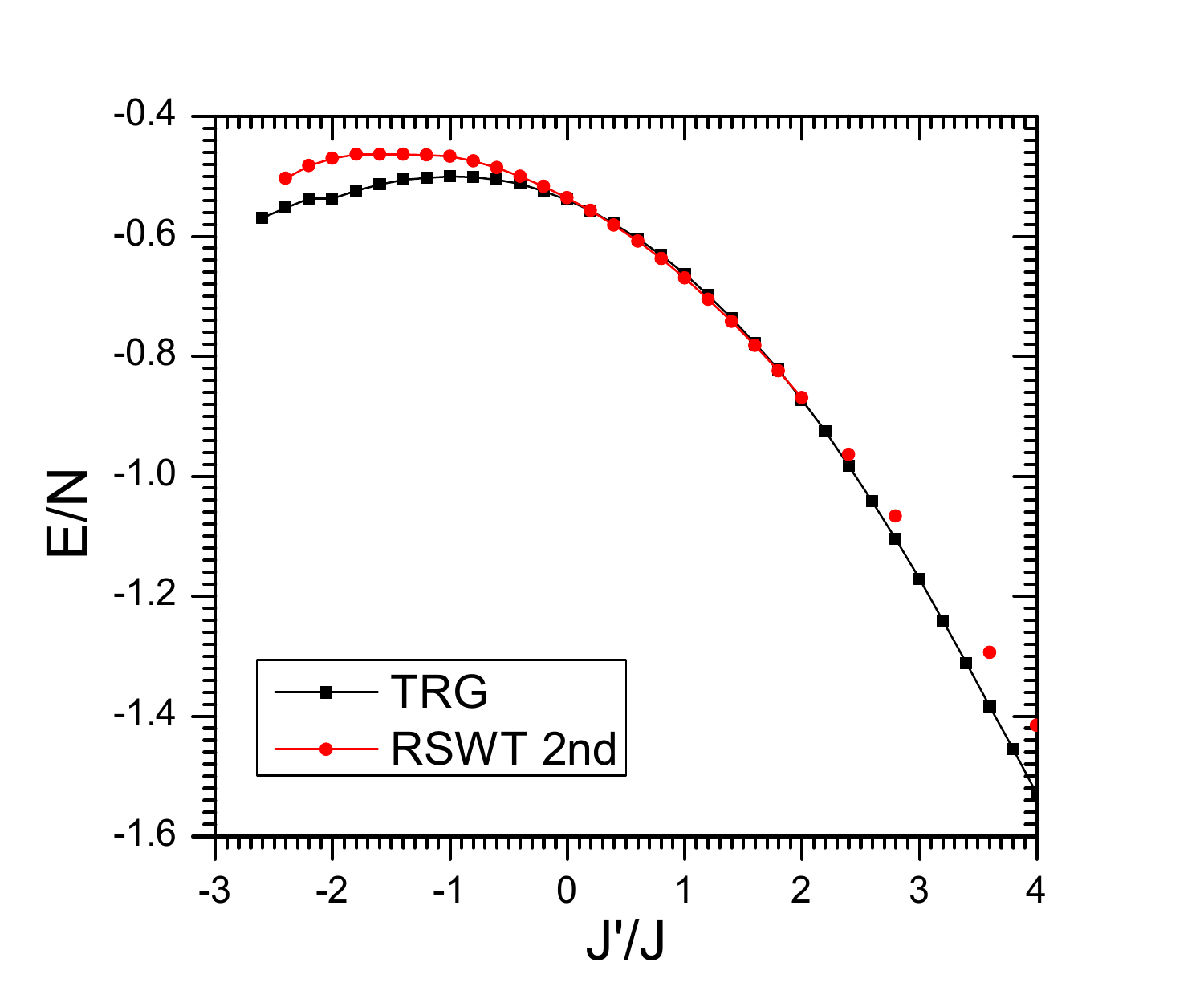,angle=0,width=10cm}}
\caption{Compare the ground state energy per site obtained from our
TRG method and the RSWT. } \label{energy}
\end{figure}

  Since the system is frustrated in the regime of $J'/J <0$, our method based on TPS and TRG will be more reliable than the others. Despite that, we still compare with the results from the other methods.  It was shown that classically there is a second order phase transition from the N\'{e}el phase to a helical phase at $J'/J \simeq -1/3$ based on renormalized spin wave theory (RSWT) \cite{RSWT96} or exact diagonalization (ED) \cite{CCM00}. Furthermore, it was claimed in \cite{CCM00} that the critical point shifts to $J'/J \simeq -1.35$ by using the coupled cluster method (CCM) to take the quantum fluctuations into account.  However, we do not see such a shift in our results.  Instead, we evaluate the ground state energy by TRG method and find the value is very close to the one obtained by the method of RSWT for a helical phase, especially in the N\'{e}el (classical) regime. The result is shown in Fig. \ref{energy}. We see that for $J'/J<-0.5$  the agreement in energy between ours and RSTW's starts to deviate, and our translationally invariant ground state is energetically more favored than the non-translationally invariant helical state.  Similar deviation occurs for $J'/J>2$, this again reflects the relevance of quantum effect when the system is away from the classical N\'{e}el ordered regime.

\begin{figure}[ht]
\center{\epsfig{figure=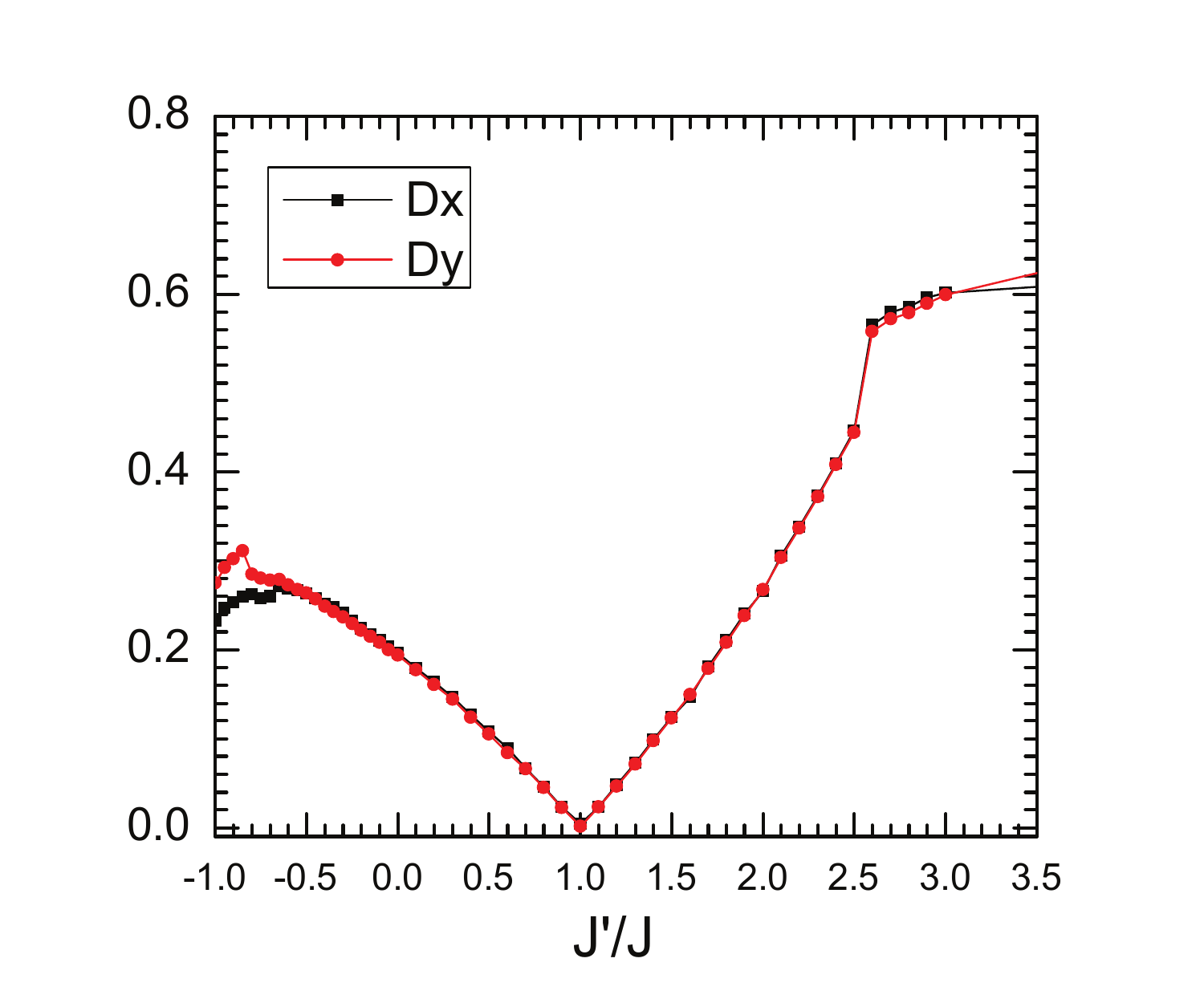,angle=0,width=10cm}}
\caption{Dimerization $D_x$ and $D_y$ vs. $J'/J$. } \label{dxdy}
\end{figure}

\begin{figure}[ht]
\center{\epsfig{figure=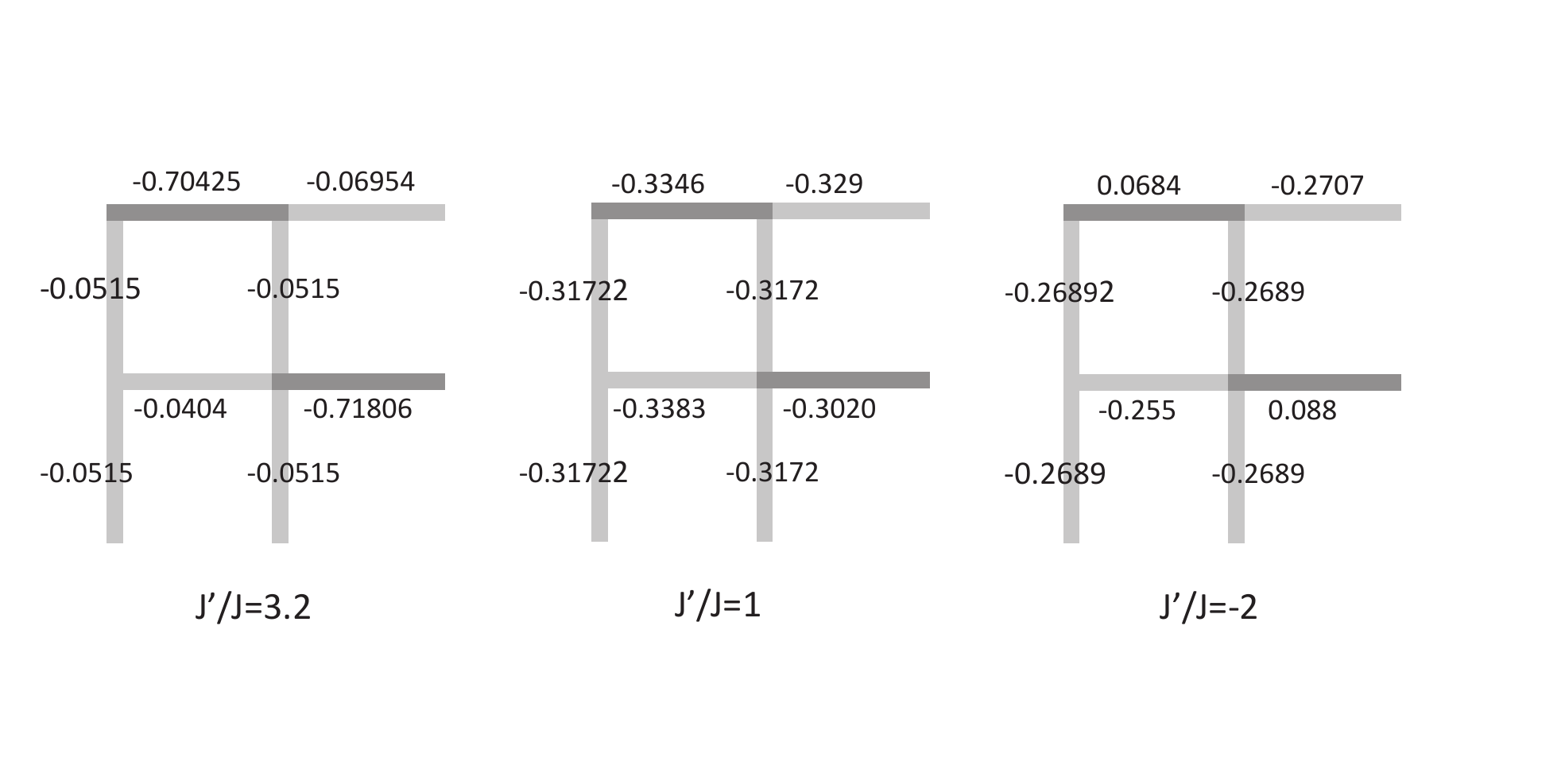,angle=0,width=12cm}}
\caption{Visualization of Spin-Spin correlation of neighboring sites
for $J'/J=3.2$ (disordered dimer phase), $J'/J=1$ (N\'{e}el phase)
and $J'/J=-2$(possibly topologically ordered phase). } \label{bond}
\end{figure}

  For curiosity, we also calculate the dimerization $D_x$ and $D_y$ defined in \eq{Dx} and \eq{Dy} though they are not the order parameters of our dimerized model.  The results is shown in Fig. \ref{dxdy}.  We see that $D_x=D_y=0$ at $J'/J=1$ as expected, otherwise they are nonzero and reach some constant values at the disordered (dimer) regime. To visualize the distribution of the dimer strength, in Fig. \ref{bond} we explicitly show the values of the spin-spin correlation of the neighboring sites for three different $J'/J$ values.

\begin{figure}[ht]
\center{\epsfig{figure=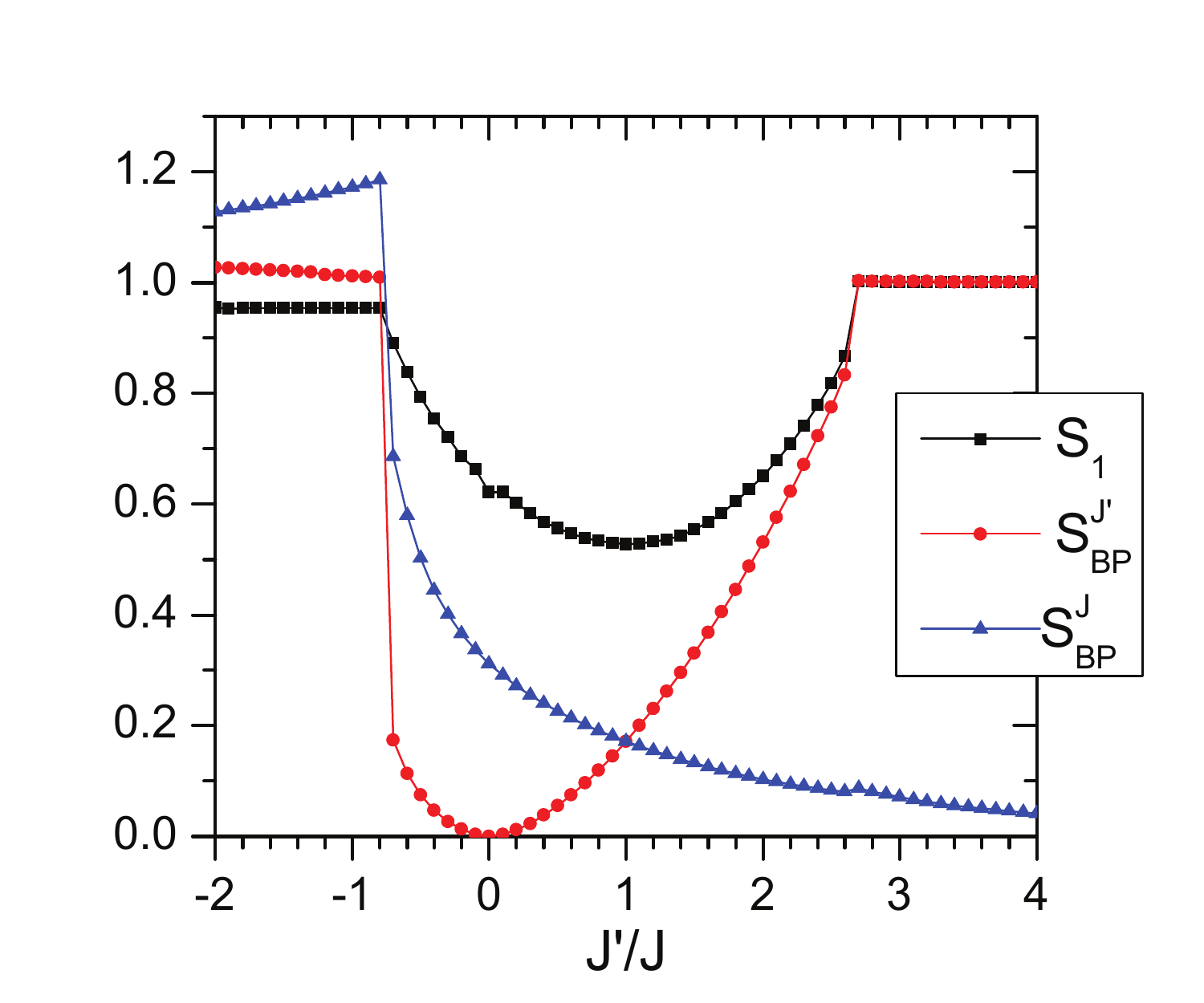,angle=0,width=10cm}}
\caption{The entanglement measures $S_{BP}^{J,J'}$ and $S_1$ v.s.
$J'/J$. The discontinuities in the derivative of the entanglement
measure occur right at the quantum critical points, and the
classical N\'{e}el regime has lower entanglement. } \label{entangle}
\end{figure}

   Beside using the order parameter such as  $\langle M^z_s \rangle$ in this case to characterize the quantum phase transitions, the entanglement measure can also do the job, see \cite{huang-lin} for some examples. Especially, for the TPS ansatz it is quite straightforward to evaluate the bipartite entanglement per bond $S^{\alpha}_{BP}:=-\sum_{i}\lambda_{\alpha,i}^2 \log \lambda_{\alpha,i}^2$, where $\alpha=J,J'$ labels the two different types of the bonds, and the $\lambda_{\alpha,i}$'s are the singular values obtained from the SVD of the site tensors in performing the Iterative Projection algorithm.  Similarly, one can also use TRG to obtain the one-site reduced density matrix, and then evaluate the von Neumann entropy, i.e.,  1-tangle denoted as $S_1$.  The results are summarized in Fig. \ref{entangle}. Note that the derivatives of both $S_1$ and $S_{BP}^{J,J'}$ are discontinuous at two quantum critical points. Moreover, it shows that the classical N\'{e}el regime has lower entanglement compared to the other two phases dictated by the significant quantum effect. For the disordered dimer phase with $J'/J>2.56$ quite amount of the quantum entanglement is due to the formation of the dimers, i.e., nearest-neighboring sites form the maximally entangled states (Bell states).  Once the dimer is formed, the sites on the both ends of the dimer will not be correlated with the other sites, i.e., the monogamy of the entanglement.

    The local nature of short range entanglement for the dimer phase is different from the long range entanglement for the topological phase in the regime $J'/J<-0.54$. Later, we will see that the difference between these twos will be reflected in  the degeneracy of the singular value spectrum.

\section{Topological Entanglement Entropy}

   In this section, we will evaluate the topological entanglement entropy based on TPS ansatz. We will consider first the J-J' model, and then a toric code like toy TPS state to demonstrate the power of TRG method in identifying the topological phase.

\subsection{J-J' model}
   In order to make sure that the phase in the regime $J'/J<-0.54$ is the topological phase, we evaluate the topological entanglement entropy (i.e.,  $- \gamma$) according to the prescriptions mentioned before. We also find that the entanglement entropy increases linearly with the block boundary size, this gives a consistent check for our numerical codes.

 \begin{figure}[ht]
\center{\epsfig{figure=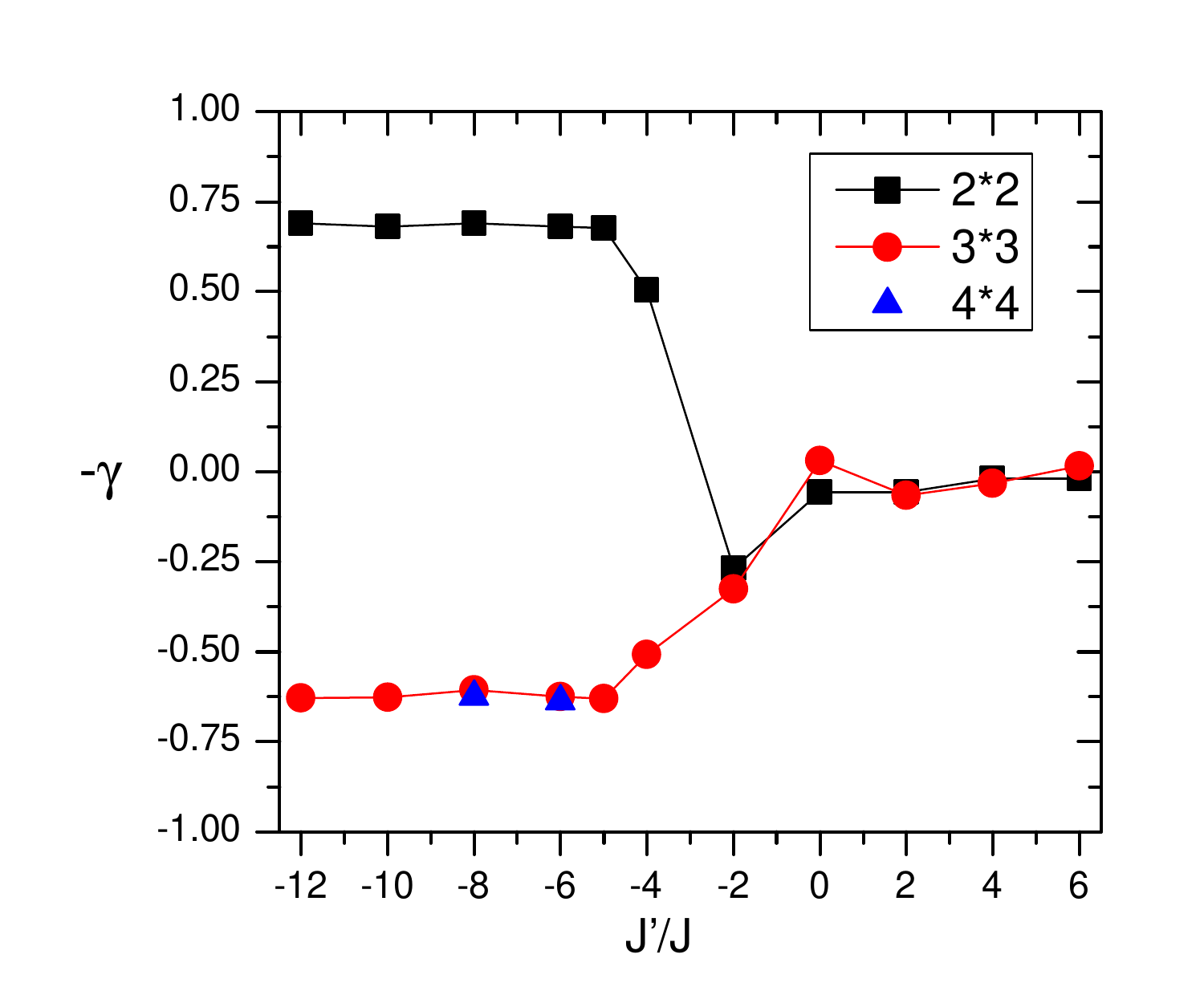,angle=0,width=8cm}} \caption{
Topological entanglement entropy vs. $J'/J$ for the 2 by 2 block (square), the 3 by 3 block (circle) and the 4 by 4 block (triangle) for J-J' model.}
\label{JJtopo}
\end{figure}

   Our result is shown in Fig. \ref{JJtopo}. The one for 2 by 2 block is evaluated by Kitaev-Preskill's scheme, and the one for 3 by 3 block by Levin-Wen's scheme. For larger blocks, it requires far more power of computation in TRG for evaluating the block entanglement entropy. For example, if we would like to evaluate the entanglement entropy for the 4 by 4 block,  then the number of entries for the reduced density matrix will be about $2^{16}/2^9=128$ times of the one for 3 by 3 block.  That means it will take about 100 times of the computing time than for the 3 by 3 block, and it is beyond our computational capacity. This is the disadvantage when trying to identify the topological phase with the topological entanglement entropy numerically.  Despite that, we still spent months to evaluate the topological entanglement entropy for the 4 by 4 block at $J'/J=-6, -8$ as shown in Fig. \ref{JJtopo}.  The values of these two data points are quite consistent with the ones for the 3 by 3 block. More closely, for the 3 by 3 block $\gamma=0.6248$  at $J'/J=-6$ and $\gamma=0.6057$  at $J'/J=-6$, and for the 4 by 4 block  $\gamma=0.6276$  at $J'/J=-6$ and $\gamma=0.6111$  at $J'/J=-6$, respectively.

   Our result shows that $\gamma$ is consistent with zero for the non-topological phases (N\'{e}el and dimer phases). However, we find that $\gamma$ is negative and unphysical in the topological phase for 2 by 2 block. This could be due to the small boundary size effect from the next sub-leading term of  ${\cal O}(L^{-\nu})$ with $\nu>0$.  On the other hand, for the 3 by 3 block we expect the small boundary size effect could be suppressed so that $\gamma$ in the topological phase should be positive.  We find that this is indeed the case, and this is also confirmed by the two data points for the 4 by 4 block.  In Fig. \ref{JJtopo},  there is a long crossover for $\gamma$ as we vary $J'/J$ between $0$ and $-5$. Then, $\gamma$ approaches $0.6$ around  $J'/J=-5$ and then remains at that value for more negative $J'/J$ as expected from its topological nature. Naively, we expect a quantum critical point separating topological and non-topological phase, and the crossover could be due to the short range entanglement in our TPS solution. We expect that it should be removed by suitable quantum state renormalization.

   Though our result is subjected to the errors from the small size effect and also from the approximation in our numerical method of Iterative Projection and TRG, it does indicate a nonzero $\gamma$ and thus a topological phase in the regime with $J'/J<-0.54$. We may expect the value of $\gamma$ will be improved if the above errors could be reduced.

\subsection{Toric code like state}

Due to the limitation of our computation power, it is hard to
evaluate the topological entanglement entropy to high precision by
increasing the bond dimension $\chi$ in the TPS ansatz for the J-J'
model. Instead,  we now consider a one-parameter toric code like TPS
state proposed in  \cite{2drg} with $\chi=2$. In this case we can
evaluate the topological entanglement entropy more precisely to
characterize the topological order, and demonstrate the power of TRG
method on this issue. Later on, we will compare the topological
entanglement entropy of this state with the pattern of degenerate
singular value spectrum of the same state, and demonstrate that the
latter can be also used to characterize the topological order
without invoking heavy computation capacity. Though this
one-parameter state has no relation to the J-J' model, however, it
yields the supporting evidence for the connection between
topological order and degenerate singular values in the topological
phase of the J-J' model, and possibly for more other frustrated spin
systems.

\begin{figure}[ht]
\center{\epsfig{figure=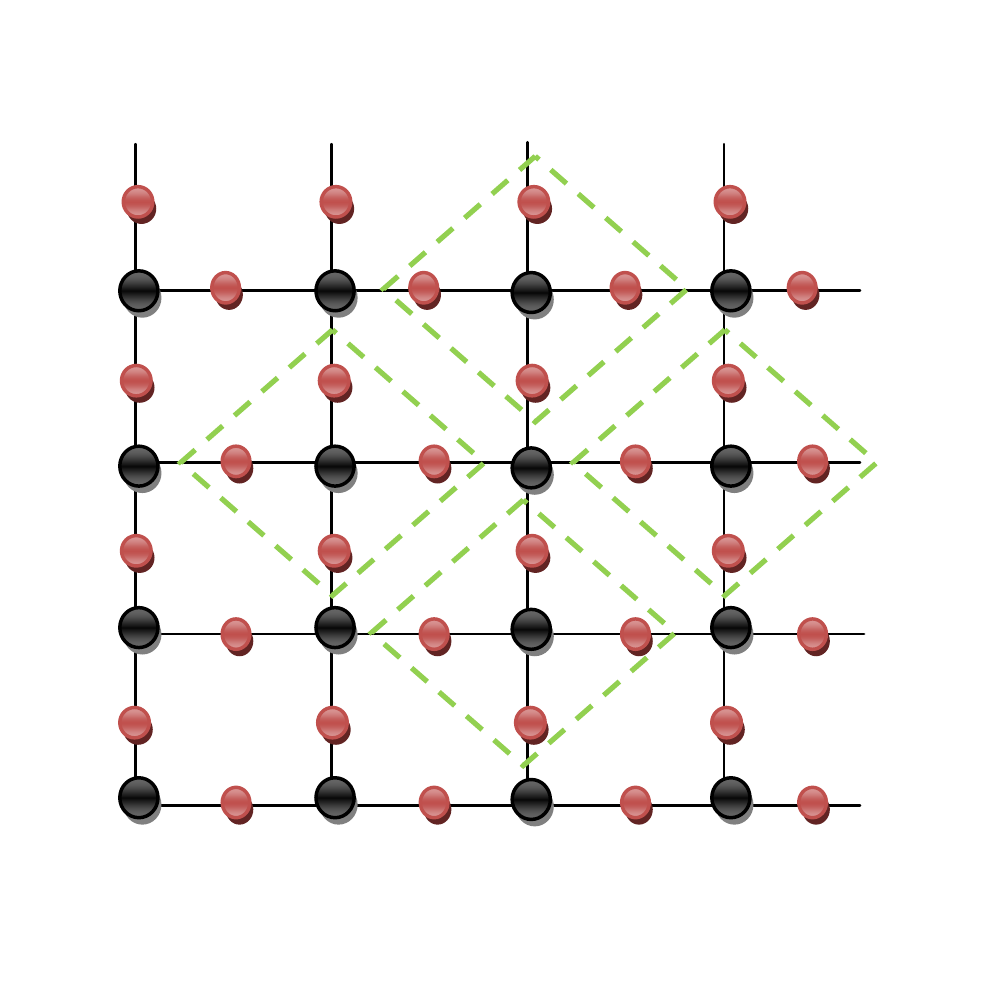,angle=0,width=6cm}} \caption{
The toric code is defined on a 2D square lattice. The red dot is a
spin-$1/2$ particle located on each link. To write down a TPS associated with the site tenor, we follow \cite{2drg} by splitting every spin-$1/2$  into two and associating each vertex with four spins,. Then, each vertex (black dot) on the lattice lives a TPS tensor with four physical indices, and four bond indices. The
(green) dashed line square is a 2 by 2 block for which we evaluate
the topological entanglement entropy for the toric code like TPS
state.} \label{toricmodel}
\end{figure}

This toy TPS state on the square lattice motivated by the toric code
model is characterized by the TPS tensor \cite{2drg} on each vertex
in Fig. \ref{toricmodel}  in the form of $T^{ijkl}_{\alpha \beta
\gamma \delta}$ with four physical indices $i,j,k,l=0,1$ and four
bond indices $\alpha,\beta,\gamma,\delta=0,1$ (i.e., bond dimension
$\chi=2$). Its entries are given by \be\label{toricT}
T^{ijkl}_{ijkl} = \left\{ \ba{cc}
g^{i+j+k+l}\;,  & \mbox{if} \quad  i+j+k+l=0 \quad \mbox{mod} \quad  2\;,   \\
 0\;,  &  \mbox{otherwise}.
\ea \right. \ee

The parameter $g$ is used to tune the property of this state from
the topological phase to the trivial phase. Note that this state is
not solved from any model Hamiltonian but is devised on purpose to
demonstrate the phase transition between topological and trivial
phases.  For $g=1$, it reduces to the ground state of the toric code
model with $\mathbb{Z}_2$ topological order.  For $g=0$, it reduces
to a product state of all zeros, i.e., a zero state. Therefore, as
we vary $g$, the state will go through a phase transition, and in
\cite{2drg} it showed that the quantum critical point occurs at
$g_c\simeq 0.8$ which separates the $\mathbb{Z}_2$ topological phase
from the phase of product state.

\begin{figure}[ht]
\center{\epsfig{figure=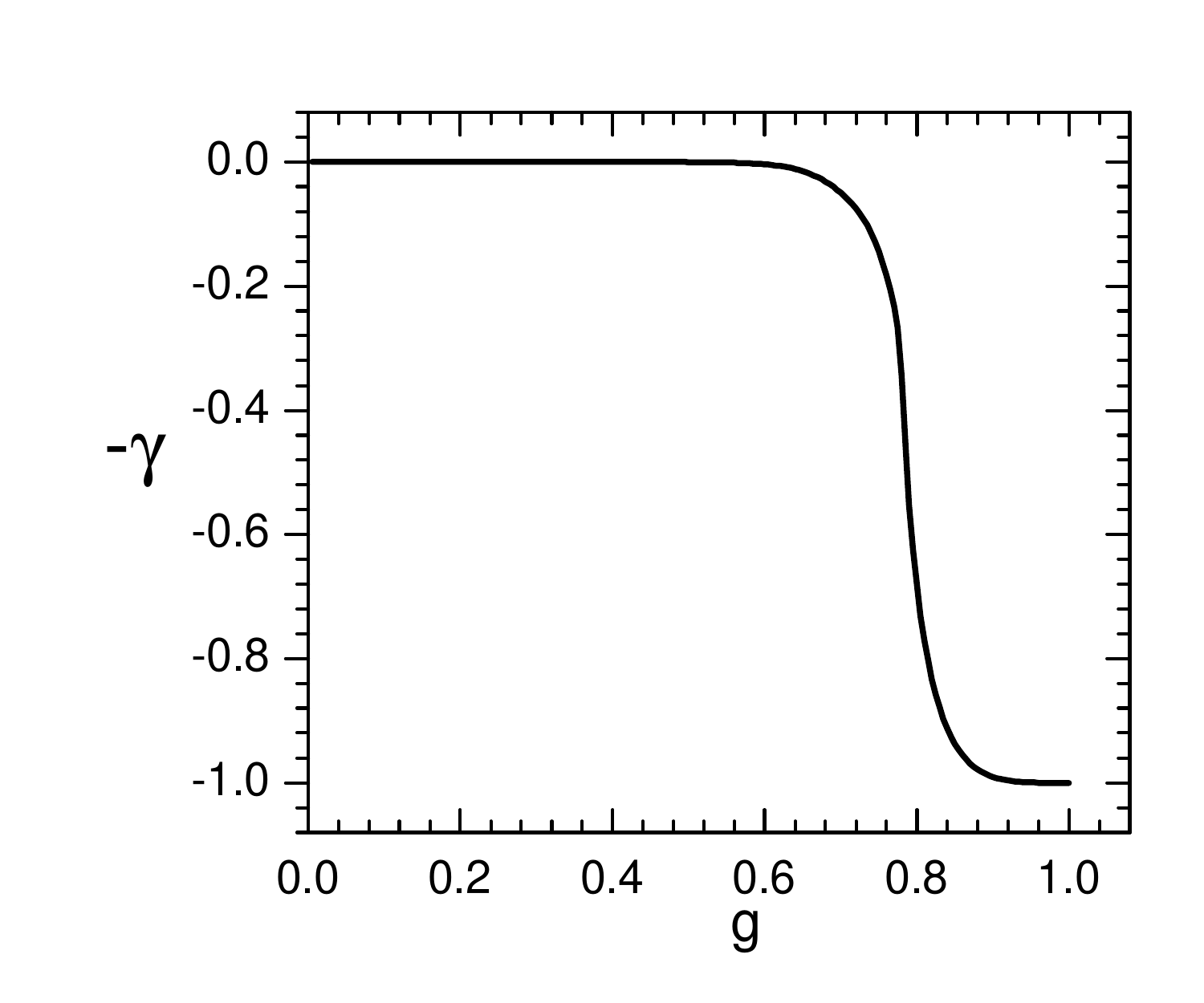,angle=0,width=8cm}} \caption{
Topological entanglement entropy from Kitaev-Preskill's scheme for the 2 by 2 block in Fig. \ref{toricmodel}.
 } \label{z2topo}
\end{figure}

With the above TPS state, we then use the Kitaev-Preskill's scheme
to numerically evaluate the topological entanglement entropy for  2
by 2 block indicated by the blue dash line in Fig. \ref{toricmodel}.
Since this state has only $\chi=2$ and thus requires not much
computation power to carry out the numerical calculation up to high
precision. Again, we verify that the entanglement entropy obeys the
area law. Our result is shown in Fig. \ref{z2topo}, and it indicates
a crossover around $g_c\simeq 0.8$ separating the trivial phase with
$\gamma=0$ and the topological phase with $\gamma=1$. We may observe
a sharp jump at $g_c$ if we increase the block size, however, it is
then beyond our computation capacity.

\section {Degeneracy of Singular Value Spectrum and Topological Order}

In general, there are some ways of classifying the possible
topological phases, however, there is no specific method with
consensus in identifying the topological phase of a particular
dynamical model. Topological entanglement entropy
\cite{kiteal06,wen06} is usually thought to be good index in
characterizing the topological order, but it is difficult to
evaluate numerically to high precision due to the limitation of the
computation capacity as discussed in the previous section.

    On the other hand, the degeneracy of the entanglement spectrum has recently implemented to characterize the topological orders for some 2D quantum Hall states \cite{Haldane08} and some 1D symmetry protected topological phase \cite{Pollmann08}. For 2D TPS, it is more easier to consider the singular value spectrum, which is related to the entanglement spectrum in some way. Despite  there is no rigorous proof on its power of characterizing the 2D topological orders, for the translationally invariant states the degeneracy of the singular value spectrum  by itself indicates the existence of the long range entanglement.  Recently it is argued that the long range entanglement is the characteristics of the topological order from the point of view of the quantum information \cite{wen1d,wen2d,classifyMPS,Pollmann08}.  The tensor of the TPS encodes the information about the entanglement for the bi-partitions of the system. The singular value spectrum obtained from it  weighs different components of quantum correlation between bi-partition regions in the ground state wave function. Therefore, if there is a double degeneracy of the singular value spectrum, with the help of the translational invariance it may imply that the bi-partitions are almost in the maximally entangled states. It is the long range entanglement since it involves the the collective coherences of the spins on half of the space to be strongly correlated with all the spins in the other half.

    By considering the frustrated regime of the J-J' model, we demonstrate the above picture is on the right track to uncover the topological nature of the long range entanglement. It needs more study on both numerical and analytical sides to solidify the above connection. From our on-going work \cite{linhuang1}, we find more positive evidences.

 \subsection{Some examples}
    Instead of providing the proof of the above picture, we give examples. A well-studied one is the symmetry-protected Haldane phase, which is described by the Afflect-Kennedy-Lieb-Tasaki (AKLT) state. The AKLT state can be put into the form of the MPS represented by the matrices $\{A_i\}=\{\sigma^z,\sqrt{2}\sigma^+,-\sqrt{2}\sigma^-\}$.    By merging two matrices and then performing SVD, we find that the singular values are doubly degenerate. This is also proved in more rigorous way in \cite{Pollmann08}. Since the topological phase is robust again small perturbations,  we consider the spin-1 model Hamiltonian
\begin{align}
H=J\sum_i\vec{S}_i\cdot \vec{S}_{i+1}+U_{zz}(S_i^z)^2\;.
\end{align}
At the large $U_{zz}$, there is a trivial phase where all the spins are in the eigenstates of $S^z$. On the other hand, for small $U_{zz}$ the system is in the gapped Haldane phase, which is related to the AKLT state by local unitary transformation. We use the iTEBD method to solve the ground states in the MPS ansatz and then find the singular value spectrum. We find that the phase transition between Haldane phase and the trivial phase is characterized by the degeneracy of the singular value (entanglement) spectrum.

The relation between the degeneracy of the singular value spectrum
and the topological order for the 2D spin system is less explored.
Here we consider the simplest $\mathbb{Z}_2$ toric code model
\cite{toric}, and the explicit TPS form of its ground state is
represented by the tensor $T^{ijkl}_{\alpha \beta \gamma \delta
,\mathbb{Z}_2}$ as discussed in the previous section. We can find
the singular value spectrum by merging the tensors and then
performing SVD. Again, we find that four kinds of the singular
value spectrum are all two-fold degenerate.

 \subsection {Topological order of J-J' model by singular value spectrum and its robustness}

\begin{figure}[ht]
\center{\epsfig{figure=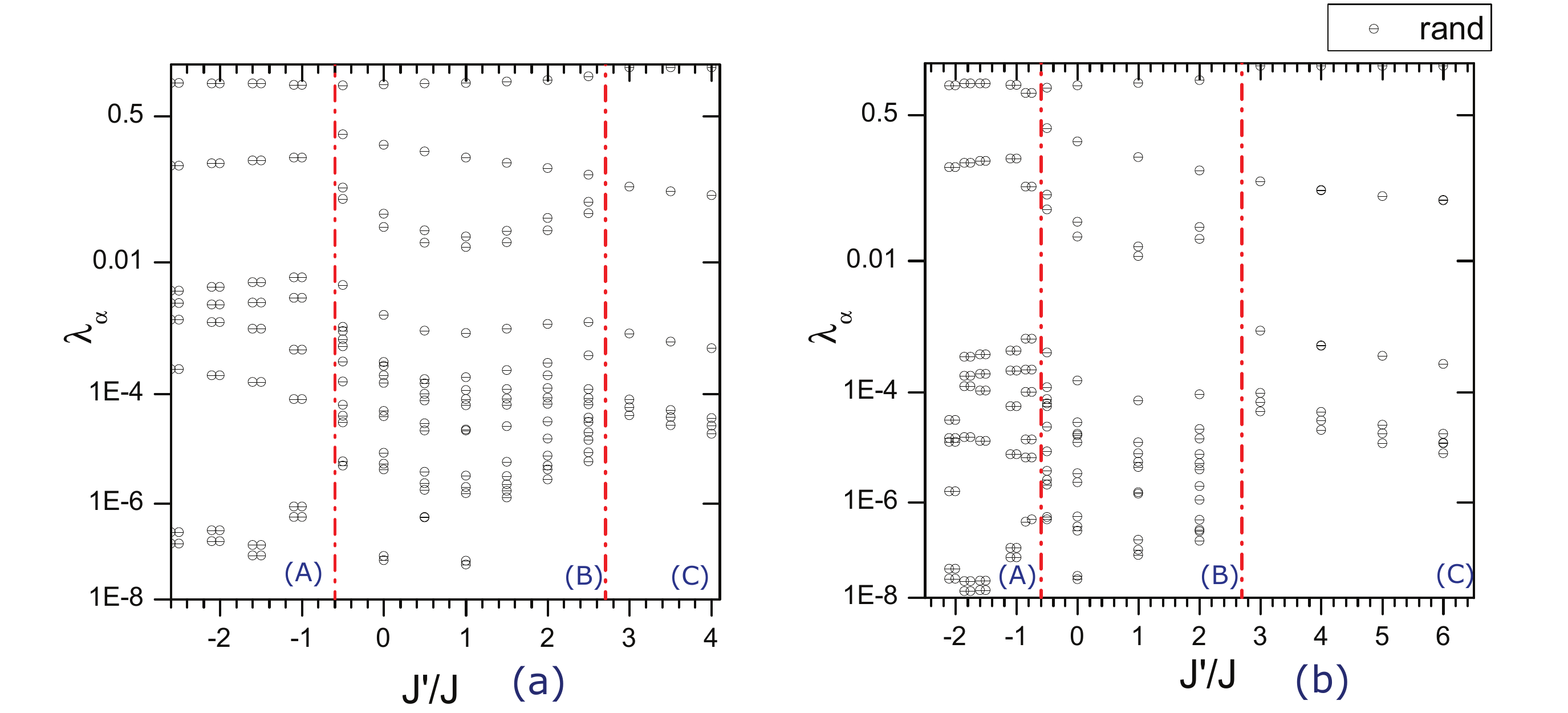,angle=0,width=16cm}} \caption{(a)
Singular value spectrum $\lambda_{\alpha}$ per bond of the J-J' model v.s. the coupling ratio $J'/J$. Each small circle with a middle bar represents a specific singular value. The double degeneracy of the singular values is denoted by the two closely touched neighboring circles. Our results show that almost all the singular values are doubly degenerate.  (b) Singular value spectrum of the J-J' model subjected to the random perturbation at each step of Iterative Projection method. We see that the pattern of singular spectrum changes but the its double degeneracy in the frustrated regime remains.}
\label{normal}
\end{figure}

The singular value spectrum for each bond in the J-J' model from the
TPS of our numerical ground state solution is given in Fig.
\ref{normal}(a). Each small circle with a middle bar represents a
specific singular value of the singular value spectrum. Note that
the singular value spectrum for the  $J$-  and $J'$-bond may show different degeneracy patterns, however, if the patterns are similar we will only show one of them.  We see that the singular values are doubly degenerate for the regime $J'/J>-0.54$ in which the  N\'{e}el order also vanishes. Therefore,
$J'/J=-0.54$ should be the critical point separating the topological
and non-topological phases, along with the evidence of nonzero
topological entanglement entropy discussed previously. On the other hand, for the dimmer phase in the regime with $J'/J>2.56$, we find the singular values for the $J'$-bond are
doubly degenerate, but others aren't.  This reflects the peculiar feature of the strong dimmer bond but not the topological order because the topologically entanglement entropy is zero in this phase. Moreover, our results show that not just the dominant singular value but almost all the singular values for that bond are also doubly degenerate.

    Recall that the precision for the topological entanglement entropy and the sharp quantum transition point is hard to achieve due to the limitation on the computation capacity. However, with the same capacity we can obtain the degenerate singular value spectrum quite easily with a sharp quantum transition point.  In contrast with the topological entanglement entropy, this is the advantage to use the degenerate singular value spectrum to characterize the topological order numerically with TPS ansatz.

    Since the topological order should be robust against the local perturbations, the degeneracy of the singular value spectrum should also does so if it can be used to characterize the topological orders.  We first add a random perturbation $H_{rand}$ at each step of Iterative Projection method, where the random 2-body Hamiltonian $H_{rand}$ is a real symmetric matrix with 10 independent matrix elements given by the random numbers in the interval $[0, 0.1J]$. The result is shown in Fig. \ref{normal}(b). We see that the singular value spectrum is changed but the pattern of its degeneracy remains intact.  This demonstrates the robustness of the degeneracy of the singular value spectrum, and thus the robustness of the topological order.

\begin{figure}[ht]
\center{\epsfig{figure=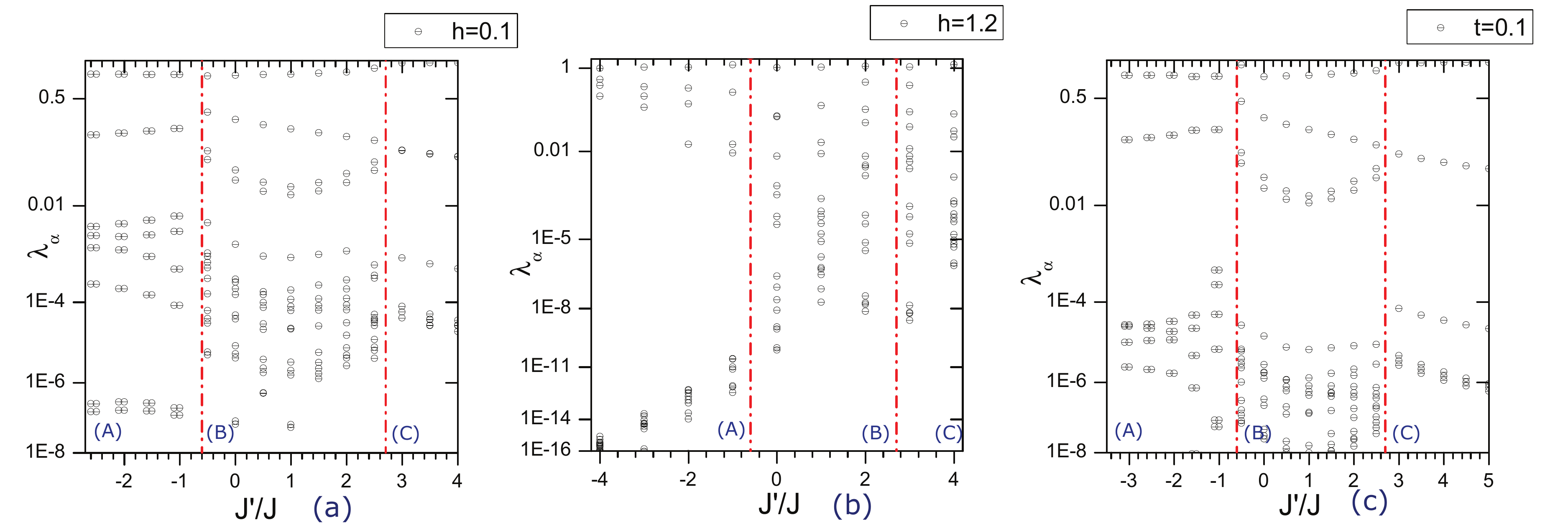,angle=0,width=18cm}}
\caption{Singular value spectrum $\lambda_{\alpha}$ per bond of the J-J' model perturbed
by $h \sum_{i,j} S_{i,j}^{z}$ with (a) $h=0.1$, (b) h=$1.2$
(critical value), and (c) perturbed by $t \sum _{i,j}(-1)^j
\vec{S}_{i,j}\cdot \vec{S}_{i,j+1}$ with $t=0.1$.
If the perturbation is too large such as in the case (b), the degeneracy will then be lifted.
These perturbations break various symmetries of the original J-J's model. Despite that, the double degeneracy of the singular spectrum remains. This demonstrates that the degeneracy pattern of the singular value spectrum is robust even under the symmetry-breaking perturbations. This feature is different from the 1D symmetry-protected topological order.} \label{addh}
\end{figure}

  To further check the robustness, we add a perturbation of the form $h \sum_{i,j} S_{i,j}^{z}$ which breaks the spin SO(3) symmetry and the time-reversal symmetry. For small $h$, the above perturbation does not destroy the double degeneracy of the singular value spectrum. Once $h$ is larger than the critical value $1.2$, the double degeneracy will be lost. The results for $h=0.1$ and $h=1.2$ are shown in Fig. \ref{addh}(a) and (b), respectively. On the other hand, in Fig. \ref{addh}(c) we show the result by adding a perturbation of the form $t \sum _{i,j}(-1)^j \vec{S}_{i,j}\cdot \vec{S}_{i,j+1}$ which breaks the translational symmetry in the direction perpendicular to the dimers.  Again, we see that the perturbation with $t=0.1$ cannot destroy the double degeneracy.

  With above generic types of perturbations, we see that the degeneracy pattern of the singular value spectrum  is quite robust, this reflects the topological nature of the phase.

 \subsection {Remove short range entanglement}

   As described previously, quantum state renormalization consists of two steps: the first step is to optimize the local unitary transformation to remove the short range entanglement, and the second step is to coarse grain the site tensors according to the TRG like method while keeping the physical bonds.  After successive steps of quantum state renormalization, the TPS will flow to a fixed point with most of short range entanglement being removed. Therefore, one would expect the fixed-point TPS will be the representative of the universal class of topologically ordered states. This quantum informational way of classifying the topological orders are recently proposed in \cite{wen1d,wen2d,classifyMPS,Pollmann08,2drg}.

\begin{figure}[ht]
\center{\epsfig{figure=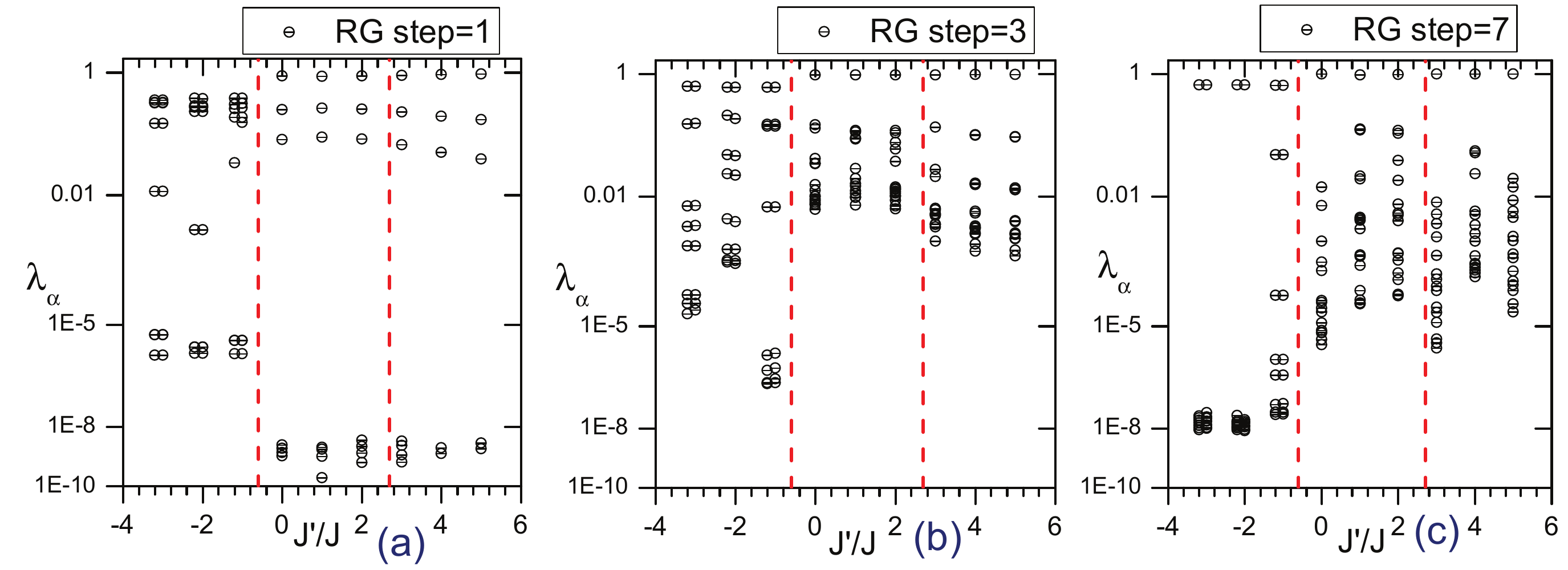,angle=0,width=18cm}} \caption{ The
singular value spectrum of J-J' model under quantum state RG flow. The number of
the RG steps is labelled accordingly. In (a), only one step of quantum state RG is performed, which is not much different from (b) where three steps of RG is done. However, when seven step of RG is performed as in (c), a gap between dominant and sub-dominant singular values develops. This gapped pattern remains the same even more RG steps are performed. This implies that a RG fixed point with long range entanglement dominated is reached. The gap also indicates that the short range entanglement is removed by the local operations of quantum state RG.} \label{jjrg}
\end{figure}

   We then would like to study how the singular value spectrum evolves under the quantum state renormalization group (RG) flow procedure. Since we conjecture that the degeneracy pattern of the singular value spectrum encodes the long range entanglement, we would expect it is robust under the quantum state renormalization, which only removes the short range entanglement.  This is indeed the case for the J-J' model and the result is shown in Fig. \ref{jjrg}. Moreover, we see that under the quantum state RG flow, the number of the dominant singular values decreases in the topological phase. We can think of the extreme case that there is only one dominant doubly degenerate singular value with all others being suppressed, then the whole lattice is in the GHZ state, i.e., maximal long range entanglement. So, the suppression of the singular values except the dominant ones corresponds to the removal of the short range entanglement under the quantum state RG flow, and the long range entanglement will be encoded in the degeneracy of the dominant singular values of the fixed-point TPS.

\begin{figure}[ht]
\center{\epsfig{figure=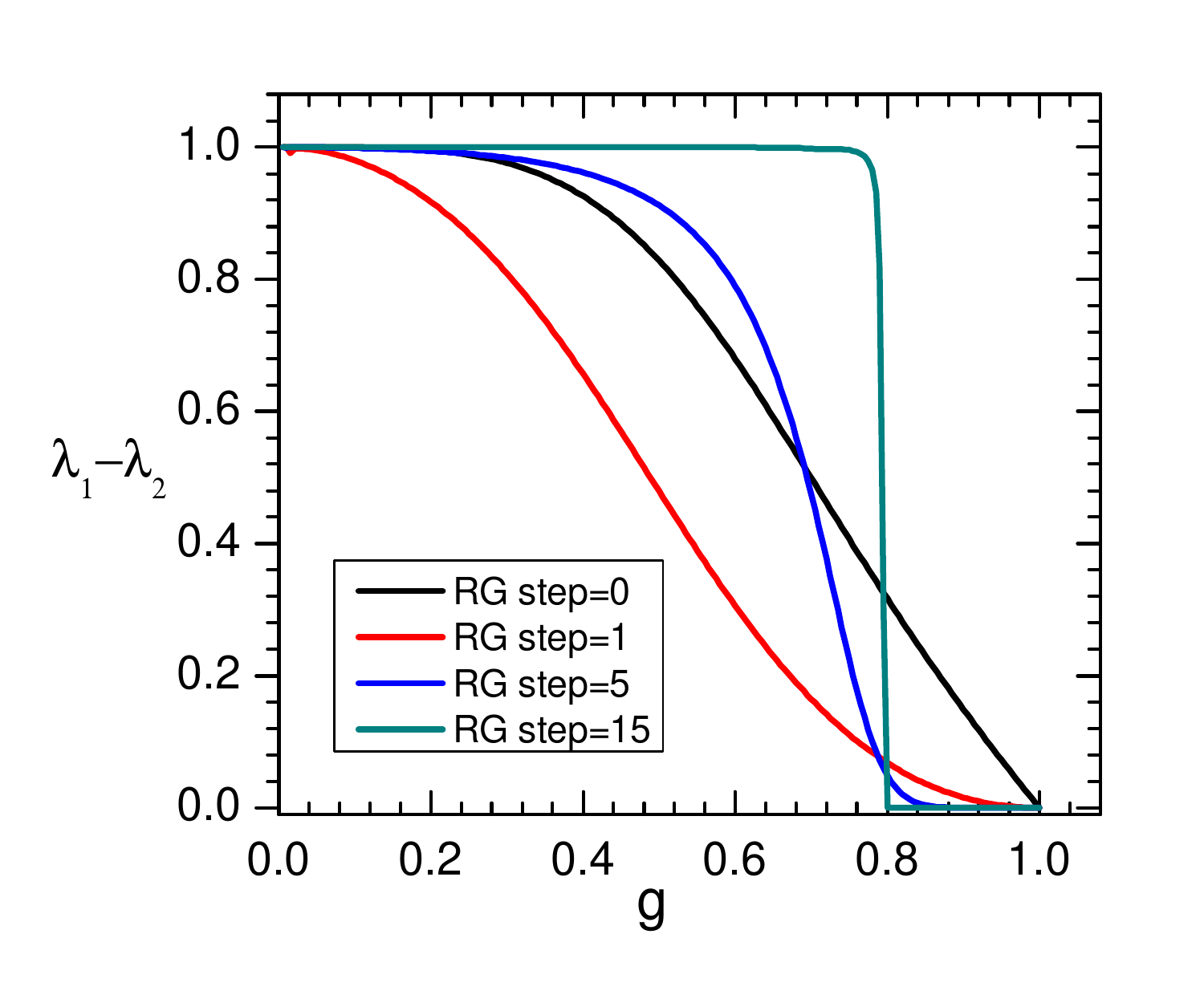,angle=0,width=8cm}} \caption{ The
difference between  the two largest singular values of toric code like model
as a function of $g$ under the renormalization flow.
 } \label{rgtoric}
\end{figure}

 Finally, to get more supporting evidences about the topological nature of the singular value spectrum under quantum state RG flow, we consider the case for the toric code like TPS state \eq{toricT} discussed before. To draw a parallel comparison with the topological entanglement entropy presented in Fig. \ref{z2topo}, we evaluate the difference between the two largest singular values, denoted as $|\lambda_1-\lambda_2|$. The result is shown in Fig. \ref{rgtoric}.

 From Fig. \ref{rgtoric}, we see that before quantum state renormalization, it seems there is no degenerate singular value spectrum
except at $g=1$. Moreover, the degeneracy at $g=1$ is 2-fold, which may
reflect the underlying $\mathbb{Z}_2$ symmetry apart from the long range entanglement.  This is in sharp contrast to the crossover behavior of topological entanglement entropy around $g\simeq 0.8$ as shown in Fig. \ref{z2topo}.
However, after one step of quantum state renormalization, we obtain the crossover  similar to the one for topological entanglement entropy. As we perform more steps of quantum state renormalization, the crossover become sharper and sharper around $g_c=0.8$. It finally approaches to a sharp quantum critical point separating the topological phase ($g>g_c$) from the non-topological one ($g<g_c$). Similar behavior for some scale invariant quantity of the same state \eq{toricT} under quantum state RG flow is also found in \cite{2drg}.

Moreover, we can get some lesson as follows. Though the degeneracy
of the singular value spectrum may indicate the existence of the long range entanglement, the reverse statement may not be true due to the contamination of the short range entanglement. However, the hidden
long range entanglement and the underlying symmetry such as $\mathbb{Z}_2$ for the toric code ground state will emerge under the quantum state RG flow as shown in Fig. \ref{rgtoric}. This is consistent with the robustness check for the topological order of toric code done in \cite{RobustnessTP}.

 \*\\
 \section{Conclusion}

    In this paper, we identify a new topological phase for the J-J' model by explicitly calculating the phase diagram of order parameters, topological entanglement entropy and the degeneracy of the singular value spectrum. This result demonstrates that the recent quantum information inspired methods such as TPS, Iterative Projection method and TRG are powerful enough in tackling the frustrated spin systems and identifying the topological orders. Especially, we demonstrate that the concepts of short range entanglement and long range entanglement introduced in \cite{wen1d,wen2d} are very useful in providing intuitive picture for the entangled nature of topological order. It is also helpful in interpreting the numerical results based on TPS ansatz.

     We have also found an intriguing connection between the degeneracy of the singular value spectrum and topological order by explicitly solving the J-J's model as a nontrivial example.  The topological order may not always reflect in the degeneracy of the singular value spectrum due to the disturbing short range entanglement. However, as we show explicitly,  we can remove short range entanglement by quantum state renormalization so that the degeneracy of the singular value spectrum after quantum state renormalization indicates the topological order.  The implication of such a connection is far-reaching in numerical identification of the topological phase in 2D systems since the computation power of obtaining the singular value spectrum is far less than the one of evaluating the topological entanglement entropy.

     Based on our observation, we believe that we can use the degeneracy of the singular value spectrum based on TPS ansatz to identify the topological phases of other interesting frustrated 2D systems. We will report more works along this direction in the near future.

\section*{Acknowledgements}
  We thank H.Y. Chen, Z. C. Gu and A. Sandvik for helpful discussions. We also thank Yu-Cheng Lin for the generous support on computer facility. This work is also supported by Taiwan's NSC grant  97-2112-M-003-003-MY3. We also thank the support of NCTS.

 \end{document}